# MECHANISM OF THE DARK MATTER AND CONDENSED BUBBLE OBJECTS FORMATION IN THE MODEL OF EXTENDED SPACE

V.A.Andreev [1], D.Yu.Tsipenyuk [2]

1- Lebedev Physics Institute of Russian Academy of Sciences,
Moscow, Russia; e-mail: andrvlad@yandex.ru
2- Prochorov General Physics Institute of Russian Academy of Sciences,
Moscow, Russia; e-mail:tsip@kapella.gpi.ru

## Abstract.

Within the framework of Extended Space Model (ESM) the processes connected to birth of photons in a gravitational field are studied. These photons have a nonzero mass. It can be both positive, and negative, and photon's energy and strength of the gravitational field determine its absolute value.

It is shown that in ESM model formation of bubble gravitational structures is possible. In the frame of ESM one can obtain the follow physical picture. Bubble gravitational objects have a halo formed by dark matter generated by photons with a positive mass. The photons with a negative mass are throw away in free deep space and create there antigravitating vacuum with negative pressure.

The comparison ESM bubble structures with similar objects of a type "gravastar", existing in a General Theory of Relativity (GR) is made.

## Introduction.

In paper [1] within the framework of Extended Space Model (ESM), developed by the authors [2,3], the gravitational effects, such as a red shift, deflection of light, radar echo and perihelion precession of Mercury were considered. It was shown that the ESM predictions for values of these effects as a first approximation coincide with predictions of General Theory of Relativity (GR).

During last years astronomers detect new phenomena, which are not described in traditional point of view on universe structure based on GR. Essences of these phenomena are reduced to following statements [4,5,6,7,8].
1) The main part of Universe mass (more than 0.9) makes dark matter and dark energy, which associate with space vacuum.
2) This dark substance does not emit electromagnetic radiation and does not interact with him, but shows gravitational properties.
3) The space vacuum has negative pressure, or, in other words, shows properties of an antigravitation, which determines dynamics of Universe extension.

In present work we develop an approach to explanation of above-mentioned phenomena based on ESM. This model is a generalization of Einstein theory of relativity at (1+4)-dimensional space, where the 5-th coordinate is the interval.
Movement along additional 5-th coordinate corresponds to changing of particles rest-mass.

This is the case when a photon, get into an exterior field, gains a nonzero mass. Moreover, this mass can have both positive and negative sign. Pair of photons which born in an exterior field consists from one photon with positive mass, and the other one has negative mass. According to ESM the dark matter is formed by massive photons. The photons with a positive mass are concentrated around massive stars and black holes. Such photons create their halo. The photons with negative mass are throw in free deep space and create there antigravitating vacuum with negative pressure. Therefore, from our point of view, the dark matter consists mainly of photons with a positive mass, and the dark energy is generated by photons with a negative mass.

Various modes that propose exist of nonzero photon mass are discussed in review [9]. Let us mark also that the possibility of existence of body with a negative mass was considered and in GR [10].

Recently attention is attracted with a new gravitational model, so-called "gravastar", or gravitational condensed star [11]. This object was offered as alternative to black holes. Such object corresponds to the solution of the Einstein equation, which outside of area occupied by masses, coincides with Schwarzchild solution. Inside of gravastar there is other, nonsingular solution, so that metric is as a whole received nonsingular.

Gravastar has a structure similar to a structure of a bubble. This bubble has a dense rigid envelope, which is under tension because of liquid substance, pressed apart it from within. Just with the help of such model, some authors try now to explain a nature of some observable objects in space.

In the present work we will show, that in the frameworks of ESM it is possible result when from a bunch of homogeneous substance will be generated the structure possessing a denser shell and the less dense center part. For this purpose we attribute to gravitation field some index of refraction $n(r)$. This index of refraction cause diminishing photon velocity and origination nonzero photon mass. In [12,13] there were considered different possibilities of attribution some refractive indexes to gravitational field.

## Photons in gravitational field.

Energy of a particle in a central field consists of two parts: potential and kinetic. In the present work we use a nonrelativistic approximation, as we are interested possibility of formation of a dark matter, and the relativistic corrections are important only for exact quantitative calculations.

Let us consider the process, in which a photon is created in some interaction of elementary particles, for example as a result of annihilation. Generally speaking in such reactions several photons are usually created, but we shall consider all these photons separately. We assume that photons are created free, massless, and only then, appearing in an external field gain a mass.

In frameworks of ESM a 5-vector [2]

$$\left(\frac{\hbar\omega}{c}; \frac{\hbar\omega}{c}\vec{k}, 0\right). \tag{1}$$

corresponds to a free photon.

The influence on it from external fields are described by rotations in extended space G(1,4). Type of a rotation and its value are determined by strength of the field and by that process, in which the particle participates, i.e. which one of the particle components varies: an energy, pulse, or mass. In this case, we consider the process, in which the energy of a particle does not vary, and only the reorganization of its interior structure happens, in particular, it's mass changed. Such rotation is described by a rotation in the plane (XS). This rotation has a form [1,2]:

$$\left(\frac{\hbar\omega}{c}; \frac{\hbar\omega}{c}, 0, 0, 0\right) \to \left(\frac{\hbar\omega}{c}; \frac{\hbar\omega}{c}\cos\psi, 0, 0, \frac{\hbar\omega}{c}\sin\psi\right) =$$
$$= \left(\frac{\hbar\omega}{c}; \frac{\hbar\omega}{cn}, 0, 0, \frac{\hbar\omega}{cn}\sqrt{n^2-1}\right). \tag{2}$$

The rotation at angle $\psi$ is determined by strength of an exterior field. Turn value and, accordingly, the parameters of a massive photon are set by the place, in which this photon is born. In this case, the photon gains a mass:

$$m = \frac{\hbar\omega}{c^2}\sin\psi = \frac{\hbar\omega}{c^2 n}. \qquad (3)$$

and velocity:

$$v = c\cdot\cos\psi = \frac{c}{n}. \qquad (4)$$

We assume that the gravitational field, in which photons are created, is described by Schwarzchild solution and the refraction index n(r) [12]:

$$n(r) = (g_{00})^{-1} = \left(1 - \frac{r_g}{r}\right)^{-1} \approx 1 + \frac{r_g}{r} = 1 + \frac{2\gamma M}{rc^2}. \qquad (5)$$

corresponds to it.

In the nonrelativistic approximation, potential U and kinetic T energies of the particle with parameters (3,4) have the form:

$$T = \frac{1}{2}mv^2 = \frac{\hbar\omega\sqrt{n^2-1}}{2n^3}; \quad U = -\frac{\gamma Mm}{r} = -\frac{\gamma M\hbar\omega\sqrt{n^2-1}}{c^2 nr}. \qquad (6)$$

We are interested the situations when the total energy of a particle E= T + U is negative. Just such particles are kept by a potential and do not leave at infinity. Let us find the energy E:

$$E = \frac{\hbar\omega\sqrt{n^2-1}}{2n^3} - \frac{\gamma M\hbar\omega\sqrt{n^2-1}}{c^2 nr} = \frac{\hbar\omega\sqrt{n^2-1}}{2n}\left(\frac{1}{n^2} - n + 1\right). \qquad (7)$$

The condition of particles trapping is reduced to an inequality:

$$n^3 - n^2 - 1 > 0. \qquad (8)$$

It is fulfilled under condition of n> 1.47, or

$$r_{max} < 4.26\frac{\gamma M}{c^2}. \qquad (9)$$

The formula (9) gives a maximum value of a radius, for which, created particles are not kept any more by gravitational field.

This result was obtained in the supposition, that all mass M is concentrated in a center, or else, that the particle is dot. Let's consider the case when gravitating object (star) M has the form of a full-sphere with a radius R and its mass M is distributed in it with a homogeneous denseness $\rho$.

Let us consider a situation when a photon is created in a point at a distance r from a star center. Inside this radius the mass $M(r) = \frac{4}{3}\cdot\pi\rho r^3$ is contained. Energy E of such photon is determined by the formula (7) and looks like:

$$E = \frac{\hbar\omega\sqrt{n^2-1}}{2n}\left(\frac{1}{n^2} - \frac{8\gamma\pi\rho r^2}{3c^2}\right) = \frac{\hbar\omega\sqrt{n^2-1}}{2n}\left(\frac{1}{\left(1+\frac{8\gamma\pi\rho r^2}{3c^2}\right)^2} - \frac{8\gamma\pi\rho r^2}{3c^2}\right). \qquad (10)$$

The condition of photon trapping gives a value of a minimum radius, outside of which the trapping of created photons happens:

$$r_{min} = \left(\frac{1.41 \cdot c^2}{8\gamma\pi\rho}\right)^{1/2}. \qquad (11)$$

Thus, the area, inside which the created photons are kept, is located between two radiuses: $r_{min}$ and $r_{max}$:

$$\left(\frac{1.41 \cdot c^2}{8\gamma\pi\rho}\right)^{1/2} < r < 4.26\frac{\gamma M}{c^2} = 5.67\frac{\gamma\pi\rho R^3}{c^2}. \qquad (12)$$

**Characteristic space objects parameters.**

Let's consider now some characteristic space objects [14] and compare their gravitational parameters: a mass M, radius R, denseness $\rho$, gravitational radius $r_g$, and $r_{min}$, $r_{max}$:

1) Sun:

$$M \approx 2 \cdot 10^{33} \text{g}; R \approx 7 \cdot 10^{10} \text{cm}; \rho \approx 1.4\frac{\text{g}}{\text{cm}^3};$$

$$r_g \approx 3 \cdot 10^5 \text{cm}; r_{min} \approx 2.45 \cdot 10^{13} \text{cm}; r_{max} \approx 5.7 \cdot 10^5 \text{cm}.$$

2) White dwarf:

$$M \approx 2 \cdot 10^{33} \text{g}; R \approx 10^8 \text{cm}; \rho \approx 4.7 \cdot 10^8 \frac{\text{g}}{\text{cm}^3};$$

$$r_g \approx 3 \cdot 10^5 \text{cm}; r_{min} \approx 1.3 \cdot 10^9 \text{cm}; r_{max} \approx 5.7 \cdot 10^5 \text{cm}.$$

3) Neutron star (1):

$$M \approx 6 \cdot 10^{33} \text{g}; R \approx 10^6 \text{cm}; \rho \approx 1.4 \cdot 10^{15} \frac{\text{g}}{\text{cm}^3};$$

$$r_g \approx 10^6 \text{cm}; r_{min} \approx 7.7 \cdot 10^5 \text{cm}; r_{max} \approx 1.7 \cdot 10^6 \text{cm}.$$

4) Neutron star (2):

$$M \approx 6 \cdot 10^{33} \text{g}; R \approx 10^5 \text{cm}; \rho \approx 1.4 \cdot 10^{18} \frac{\text{g}}{\text{cm}^3};$$

$$r_g \approx 10^6 \text{cm}; r_{min} \approx 7.7 \cdot 10^5 \text{cm}; r_{max} \approx 1.7 \cdot 10^6 \text{cm}.$$

5) Gravastar (1):

$$M \approx 10^{35} \text{g}; R \approx 1.5 \cdot 10^7 \text{cm}; \rho \approx 7 \cdot 10^{12} \frac{\text{g}}{\text{cm}^3};$$

$$r_g \approx 1.5 \cdot 10^7 \text{cm}; r_{min} \approx 10^7 \text{cm}; r_{max} \approx 3 \cdot 10^7 \text{cm}.$$

6) Gravastar (2):

$$M \approx 10^{39} \text{g}; R \approx 10^{11} \text{cm}; \rho \approx 2.3 \cdot 10^{20} \frac{\text{g}}{\text{cm}^3};$$

$$r_g \approx 1.5 \cdot 10^{11} \text{cm}; r_{min} \approx 1.9 \cdot 10^3 \text{cm}; r_{max} \approx 2.8 \cdot 10^{10} \text{cm}.$$

For the gravitational object had area, inside which photons created would be trapped, two conditions should be fulfilled to

$$r_{min} < r_{max}, \qquad (13)$$

$$r_g < R < r_{max}. \qquad (14)$$

One can see from the above mentioned data that in case of stars such as the Sun and white dwarf the condition (13) is defaulted. For neutron stars of a type (1) conditions (13) are carried out and the condition (14) is on a limit of realization, as for such objects their physical radius R approximately coincides with gravitational $r_g$. However, as the gravitational radius was found for a mass concentrated in one point, and the mass of a neutron star is distributed inside of the star radius, thus the gravitational radius of neutron star should have a smaller value. Thus, we found that the neutron stars of type (1) can be referred to number of objects for which the conditions (13) and (14) are fulfilled.

The approximately same arguments, but with large stipulations, it is possible to take into account in case of neutron stars of a type (2) and gravastar of all two types. In all these objects, can be a shaping of a bubble structure. For purpose of better understanding this process, we will study gravastar structure.

*Gravastar structure.*

The gravastar model was offered in work [11] alternatively to black holes. This static spherically symmetrical field with the metric:

$$ds^2 = -f(r)dt^2 + \frac{dr^2}{h(r)} + r^2(d\theta + \sin^2\theta \cdot d\phi^2). \qquad (15)$$

It is supposed, that the substance represents a medium without interior tension, in which denseness $\rho(r)$ and pressure $p(r)$ are connected by the equation:

$$\frac{dp}{dr} + \frac{\rho + p}{2f} \cdot \frac{df}{dr} = 0. \qquad (16)$$

Under these conditions Einstein equations are reduced to a set of equations on factors h (r), f (r) of the metrics (15):

$$\frac{1}{r^2}\frac{d}{dr}[r(1-h)] = 8\pi\gamma\rho; \quad \frac{h}{r \cdot f}\frac{df}{dr} + \frac{1}{r^2}(h-1) = 8\pi\gamma\rho. \qquad (17)$$

To solve a set of equations (16), (17), it is necessary to impose additional connection on magnitudes of $\rho$ and $p$. In dependence from type of this connection, there are different types of the solution of the Einstein equations.

It is supposed, that all substance is concentrated in space inside a radius $r_2$. It is areas I and II on Figure 1. Therefore in the field of III substances is absent and thus we have:

$$\rho = p = 0, \quad r_2 < r. \qquad (18)$$

Thus, outside of a radius $r_2$, in the III area on Figure 1, the metric (15) has the Swartzshild metric with coefficients:

$$f(r) = h(r) = 1 - \frac{2\gamma M}{r}; \quad r \geq r_2. \qquad (19)$$

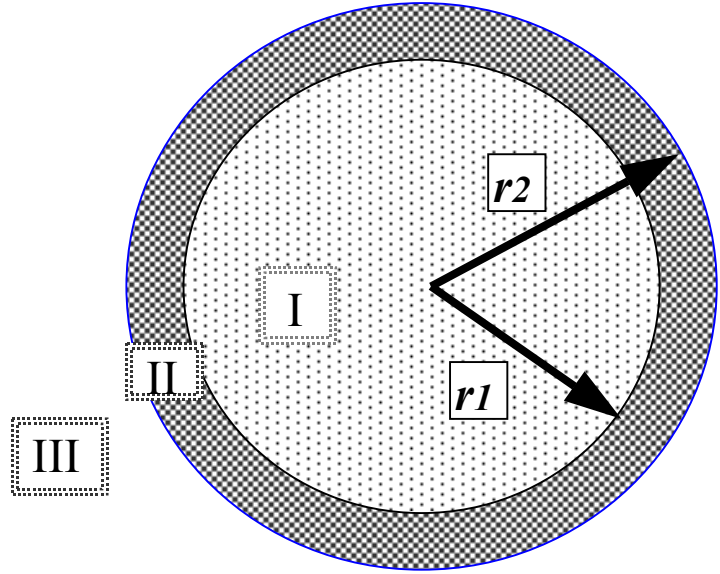

**Fig. 1
Bubbles gravastar structure.**

In such case it is supposed, that the exterior substance boundary radius is more than gravitational radius appropriate to the mass M:

$$r_g = 2\gamma M < r_2. \qquad (20)$$

Due to this in the gravastar model the metric is nonsingular and does not arise of such effects as black holes.

In interior area I of the gravastar medium follow relation is fulfilled:

$$\rho = -p; \quad 0 \leq r \leq r_1. \qquad (21)$$

Such connection between a density and pressure result to the de Sitter solution, to which there correspond coefficients:

$$f(r) = ch(r) = 1 - \frac{r^2}{R_D^2}; \quad 0 \leq r \leq r_1. \qquad (22)$$

Here $R_D$ is the radius of a curvature of the de Sitter world. By the same way as well as mass M in the Swartzshild solution, this radius arises in solutions of the Einstein equations as a constant of integration. For it the follow relation should be fulfilled:

$$R_D < r_1. \qquad (23)$$

n the field of II between radiuses $r_1$, $r_2$ follow relation is fulfilled:

$$\rho = p; \quad r_1 \leq r \leq r_2. \qquad (24)$$

In the case of realization of condition (24) equations (16) are integrated and it solution is:

$$p(r) \cdot f(r) = \text{const} = a. \qquad (25)$$

Also it appears convenient to enter a new unknown value:

$$\omega = 8\pi\gamma r^2 p. \qquad (26)$$

and to rewrite system (16), (17) in terms of variables h, w:

$$\frac{dr}{r} = \frac{dh}{1-w-h}; \quad \frac{dh}{h} = -\frac{1-w-h}{1+w-3h} \cdot \frac{dw}{w}. \qquad (27)$$

The variable w characterizes mass distribution along radius r:

$$\gamma \cdot dm(r) = w(r)dr. \qquad (28)$$

With the help of the magnitude h it is possible to calculate a distance l between points $r_1$ and $r_2$ located at the same radius:

$$l = \int_{r_1}^{r_2} dr \cdot h^{-1/2}. \qquad (29)$$

The coefficients h(r), f(r) are sewed together on the boundaries of areas $r_1$, $r_2$.

We consider gravastar model as final object, which is formed in because of processes of elementary particles disintegration. Some quantity of photons aroused under such processes. These photons are born both with positive, and with negative masses. The photons with negative masses are thrower from the object and form in deep space an area of dark energy in which is satisfied condition (21).

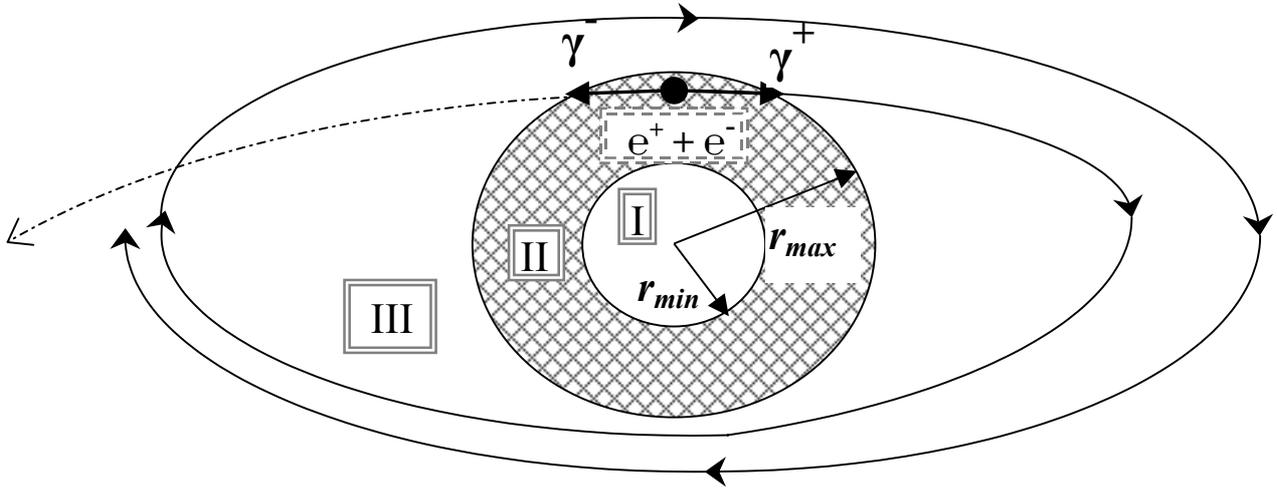

**Fig.2 Bubbles structure with a halo, originating in ESM.**

On the Figure 2 the pair of photons, originate in an outcome of an electron positron annihilation is represented. They have masses of a different sign. Photons with positive masses, in dependence from what area they were born in, or remain in a neighborhood of gravitational object and will create its halo from dark matter, or depart from it. The photons with negative masses throw in free deep space and initiate dynamics of the cosmological extension controlled by antigravitation.

In the case of defined conditions in central area of object, for $r < r_{min}$ it is possible photons trapping with negative mass and form space, in which is satisfied condition (21).

On Figures 3-6 the graphs of variables $w(r)$, $h(r)$ are reduced, where the distance $r$ is measured in units of radius $r_1$. Value of the radius $r_2$ is not fixed. Graphs forms are hardly depend from the initial conditions. It is visible, that in the case of appropriate choice of these initial conditions such relation between mass M and sizes $r_1$, $r_2$ is possible to supply in such a way that they become close to mass M and radiuses $r_{min}$, $r_{max}$ of objects considered in the previous section. The solution of the equations (17), represented on Fig.3 corresponds to case, when in a central part gravastar the area of de Sitter space is formed, as it is represented on Fig.1 with $r_1 > 0$.

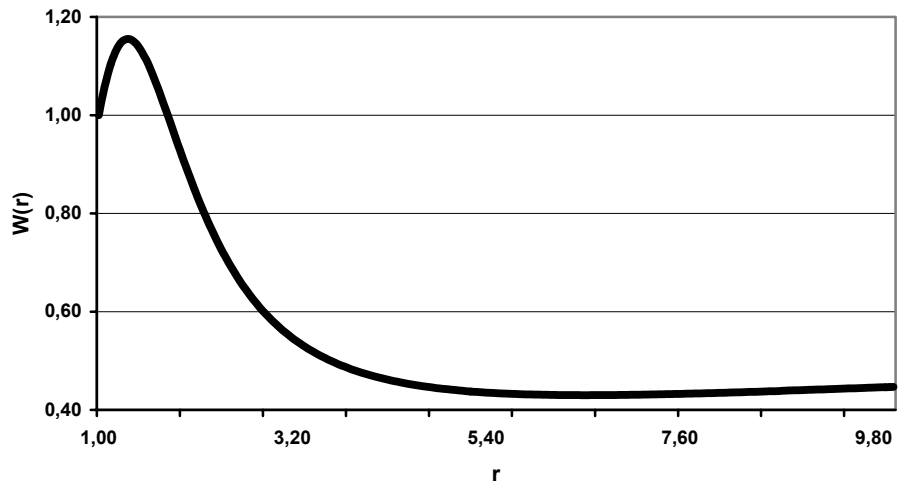

Fig.3. Variable w(r), appropriated to the initial conditions, w(1)=1, h(1)=0.99

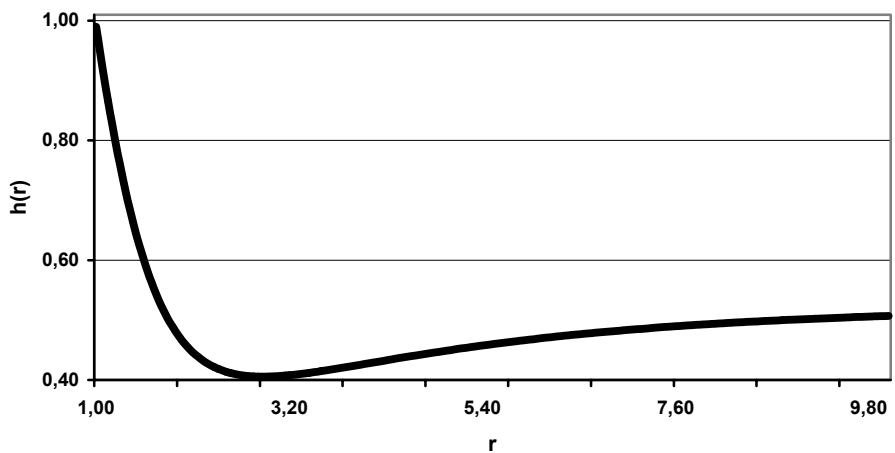

Fig.4 Variable h(r), appropriated to the initial conditions, w(1)=1, h(1)=0.99

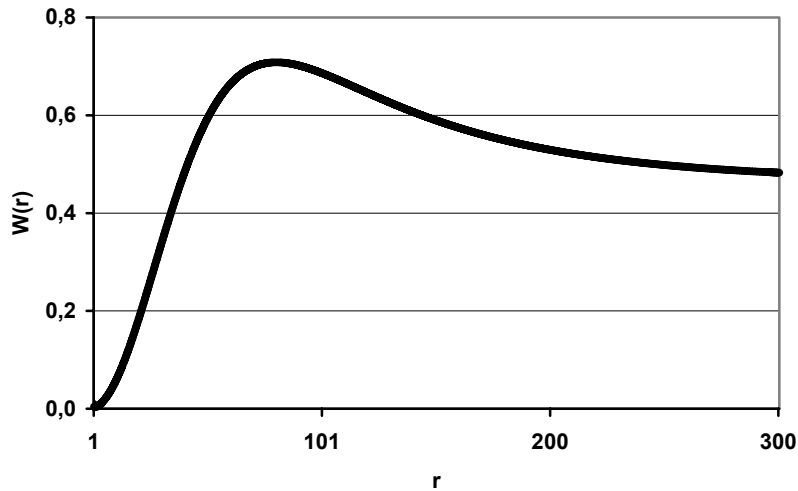

**Fig.5 Variable w(r), appropriated to the initial conditions, w(1)=0.01, h(1)=0.05**

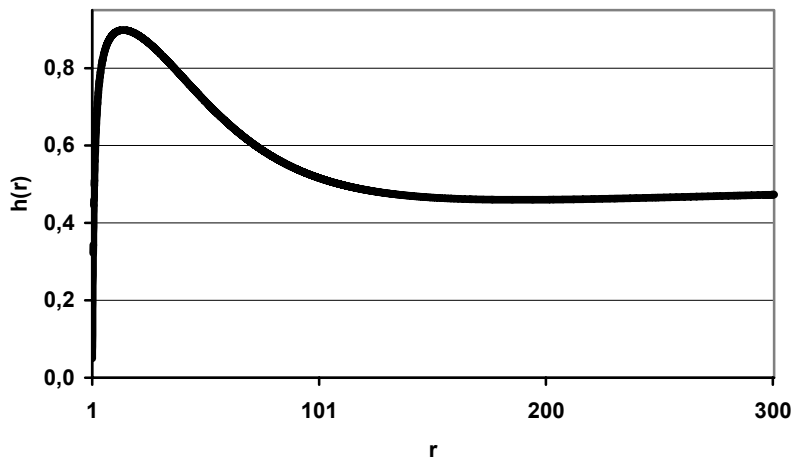

**Fig.6. Variable h(r), appropriated to the initial conditions, w(1)=0.01, h(1)=0.05**

The solution of the equations (17), represented on Fig.5, 6 corresponds to the case, when the area II can be counted directly from zero ($r_1 = 0$) and in gravastar central part will not origin areas of the de Sitter space.

Question concerning with what kind of these possibilities will be realized in case of concrete physical objects, requires additional study.

Discussion.

Thus, in ESM frameworks there is a mechanism, according to which in a neighborhood of massive gravitational objects the halo consisting of photons with positive masses can be formed. These halos we associate with a dark matter. The photons with negative masses are concentrated far from massive objects. These photons will create areas of a dark energy.

Such areas are characterized by negative pressure and exhibits properties of antigravitation. This area calls the accelerated extension of that visible part of the universe, which consists of a positive matter.

Also is shown, that the bubbles shaping structures of gravastar type it is possible. More exact reviewing of this process requires the analysis of the dynamic equations and will be made later. It is necessary also to compare these models to a model of gravitational bubbles {15}.

Bibliography.